\documentclass[12pt,showpacs,showkeys,amsmath,amssymb]{revtex4}
\usepackage{amsmath,amsfonts,amsthm,amscd,amssymb,latexsym}
\usepackage{bm}
\usepackage{dcolumn}
\usepackage{graphicx}
\usepackage{epstopdf}
\usepackage{color}
\usepackage{epsf}
\usepackage{epsfig}
\usepackage{graphicx, epic, eepic, color}

\newcommand{\beq}{\begin{equation}}
\newcommand{\eeq}{\end{equation}}

\begin{document}

\title{Critical Tidal Currents in General Relativity}

\author{Bahram \surname{Mashhoon}$^{1,2}$}
\email{mashhoonb@missouri.edu}

\affiliation{$^1$School of Astronomy, Institute for Research in Fundamental
Sciences (IPM), P. O. Box 19395-5531, Tehran, Iran\\
$^2$Department of Physics and Astronomy, University of Missouri, Columbia, Missouri 65211, USA\\
}

\date{\today}

\begin{abstract}
Relativistic tidal equations are formulated with respect to the rest frame of a central gravitational source and their solutions are studied. The existence of certain relativistic critical tidal currents are thereby elucidated. Specifically, observers that are spatially at rest in the exterior Kerr spacetime are considered in detail; in effect,  these fiducial observers define the rest frame of the Kerr source. The general tidal equations for the free motion of test particles are worked out with respect to the Kerr background. The analytic solutions of these equations are investigated and the existence of a tidal acceleration mechanism is emphasized.  
\end{abstract}


\maketitle

\section{Introduction}

In general relativity (GR), a physically reasonable and geometrically natural method to interpret the local motion of test particles relative to an observer in a gravitational field involves the introduction of a Fermi normal coordinate system along the world line of the reference observer. 

Imagine a congruence of timelike geodesics. The geodesics are neighboring, but their rates of separation may not be negligibly small compared to the speed of light $c$. We establish a Fermi coordinate system, $X^{\hat \mu} = (T, X, Y, Z)$, along the path of a reference geodesic in the congruence and consider the free motion of neighboring particles relative to the fiducial observer that follows the reference geodesic and occupies the spatial origin of the Fermi coordinate system. Limiting our attention to the two-dimensional $(T, Z)$ plane, the equation of relative motion takes the form~\cite{Chicone:2002kb} 
\begin{equation}\label{I1}
\frac{d^2Z}{dT^2} + k(T) (1-2\, \dot{Z}^2)\,Z + O(Z^2) = 0\,,
\end{equation}
where the fiducial observer is fixed at $Z = 0$, $\dot{Z} := dZ/dT$ and $k(T) = R_{TZTZ}$ is the Gaussian curvature of the $(T, Z)$ plane. Here, we use units such that $c = G = 1$, unless specified otherwise; moreover, hatted indices refer to the Fermi coordinate system.   The timelike nature of geodesic motion requires that 
\begin{equation}\label{I2}
1 - \dot{Z}^2 + k(T)\,Z^2 + O(Z^3) > 0\,;
\end{equation}
in particular, at the position of the fiducial observer ($Z = 0$), we must have $|\dot{Z}| \le 1$. Neglecting higher-order terms, Eqs.~\eqref{I1} and ~\eqref{I2} are invariant under $Z \mapsto -Z$.

When the speed of separation of neighboring geodesics is negligible, we can ignore the velocity term in Eq.~\eqref{I1} and recover the Jacobi equation in the Fermi frame. On the other hand, neglecting $O(Z^2)$ terms, the \emph{generalized Jacobi equation}~\eqref{I1} has exact solutions describing uniform motion in the Fermi system with $\dot{Z} = \pm 1/\sqrt{2}$, namely, 
\begin{equation}\label{I3}
Z_{\pm}(T) = Z_0 \pm \frac{1}{\sqrt{2}}\, (T-T_0)\,,
\end{equation} 
where $Z_0$ and $T_0$ are constants. That is, regardless of the magnitude of $k(T)$, \emph{there is a critical tidal current with speed $V_{\rm crt} = c/\sqrt{2} \approx 0.71 \,c$ such that the corresponding relative motion  in the Fermi system is approximately uniform}. In GR, a tidal current is relative motion induced by spacetime curvature. The peculiar feature of solution~\eqref{I3} of Eq.~\eqref{I1} is that these \emph{critical} tidal currents are independent of the spacetime curvature $k(T)$. This particular relativistic feature of tides in general relativity is the focus of the present paper. 

If the speed of relative motion is below the critical speed, then the nature of relative motion in Eq.~\eqref{I1} is similar to the familiar Jacobi equation. However, for \emph{relativistic} motion above the critical speed, there is deceleration for $k(T) < 0$ toward the critical speed and acceleration for $k(T) > 0$. Various aspects of this circumstance and its implications for astrophysical jets have been explored in previous work~\cite{Chicone:2002kb, Chicone:2003yv, Chicone:2004jh,  Chicone:2004pv, Chicone:2004rq, Chicone:2004ic, Chicone:2005da, Chicone:2005jj,  Mashhoon:2005pj, Kojima:2005dm, Chicone:2005vn, Chicone:2006rm, Mullari:2006sf, Perlick:2007ux, BGJ}. For recent work on general relativistic tidal effects in other contexts, see, for example, Refs.~\cite{Bini:2014rua, Iorio:2014haa, Bini:2016xqg, Puetzfeld:2018cnf, Mashhoon:2018lgk, Junior:2020yxg, Junior:2020par} and the references cited therein. 

The purpose of the present paper is to illustrate further the nature of critical  tidal currents in general relativity. In contrast to previous work regarding certain tidal acceleration phenomena involving energetic particles in astrophysics~\cite{Chicone:2002kb, Chicone:2004rq}, we assume in this paper that the reference observer is spatially static, so that the observer remains at rest in the background gravitational field. \emph{Such observers in effect define the rest frame of the background gravitational source}. Indeed, we extend here more recent work regarding jets along the axis of a Kerr field~\cite{Bini:2017uax}. In the cases considered in the present paper, nongravitational forces are needed to keep the observer fixed in space; therefore, the reference observer is generally accelerated. We further assume that the spatially static observer carries an orthonormal tetrad frame that is Fermi-Walker transported along its world line. Using these locally nonrotating basis vectors, we establish a Fermi coordinate system as described in detail in Appendix A. To study the motion of free test particles relative to the fiducial observer, we express the timelike geodesic equation in these Fermi coordinates. In this way, we can describe in an invariant way the local motion of test particles relative to the rest frame of the gravitational source. The resulting generalized Jacobi equation and the timelike condition are given by Eqs.~\eqref{A14} and~\eqref{A15} of Appendix A, respectively. In this case, critical tidal currents are described in Section II. The analytic solutions of the tidal equations  are explored in the exterior Kerr and Schwarzschild spacetimes in Sections III and IV, respectively. Section V contains a discussion of our results. Detailed calculations are relegated to Appendices B and C.

\section{Critical Tidal Currents}

The tidal equations given in Appendix A are rather complicated. To illustrate a significant feature of these equations involving critical tidal currents, let us imagine a Fermi system of coordinates $X^{\hat \mu} = (T, X, Y, Z)$ and a circumstance where motion purely along a certain Fermi coordinate direction becomes possible and the corresponding component of acceleration vanishes. Let us call this direction again $Z$ and note that 
Eqs.~\eqref{A14} and~\eqref{A15} reduce in this case to Eqs.~\eqref{I1} and~\eqref{I2}, respectively, where curvature $k$ is now \emph{constant} because the fiducial observer is at rest in space and the gravitational field is assumed to be stationary as in the Kerr spacetime (see Appendices B and C).  Ignoring higher-order terms for the moment, we can write the autonomous tidal equation as
\begin{equation}\label{I4}
 \dot{Z} \,\frac{d \dot{Z}}{dZ} + k \,(1-2\, \dot{Z}^2)\,Z = 0\,,
\end{equation}
which can be simply integrated. The result is
\begin{equation}\label{I5}
 \dot{Z}^2 = \frac{1}{2} - \left(\frac{1}{2}  - \dot{Z}_{\rm int}^2\right)\,e^{2\,k\,Z^2}\,
\end{equation}
with the initial conditions that at $T = 0$, $Z = 0$ and $\dot{Z} = \dot{Z}_{\rm int}$. Integrating Eq.~\eqref{I5}, we find
\begin{equation}\label{I6}
 \int_{0}^{Z} \left[1+\left(2\, \dot{Z}_{\rm int}^2 -1\right)\,e^{2\,k\,\zeta^2}\right]^{-1/2}\,d\zeta = \pm V_{\rm crt} \,T\,,
\end{equation}
where only positive square roots are taken into account throughout.  Here, as before,  $V_{\rm crt} = c / \sqrt{2} \approx 0.71 \,c$ is the critical speed of tidal currents; that is, for $\dot{Z}_{\rm int}^2 = V_{\rm crt}^2 = 1/2$, Eq.~\eqref{I6} implies $Z = \pm V_{\rm crt}\, T$. Clearly, this uniform motion is rather approximate; in fact, higher-order tidal terms introduce significant modifications near the boundary of the Fermi coordinate patch. Furthermore, the Fermi coordinate speed must satisfy the timelike condition, namely, $1-\dot{Z}^2 + k\,Z^2 > 0$ in accordance with Eq.~\eqref{I2}. 

For insight regarding tidal motions away from critical currents, it proves useful to  write Eq.~\eqref{I5} as
\begin{equation}\label{I7}
 \dot{Z}^2  + \left(\frac{1}{2}  - \dot{Z}_{\rm int}^2\right)\,e^{2\,k\,Z^2}= \frac{1}{2}\,
\end{equation}
and interpret this relation as an effective energy equation for a classical particle with total energy $V_{\rm crt}^2 = 1/2$ and symmetric effective potential energy $\mathcal{V}_{\rm eff}$, where
\begin{equation}\label{I8}
\mathcal{V}_{\rm eff} = (V_{\rm crt}^2 - \dot{Z}_{\rm int}^2)\,e^{2\,k\,Z^2}\,.
\end{equation}
The effective kinetic energy is always positive; therefore, tidal motion is confined to the region where $1 \ge 2 \,\mathcal{V}_{\rm eff}$. Moreover, turning points occur where 
$1 = 2 \,\mathcal{V}_{\rm eff}$. 

For $k > 0$, the absolute magnitude of the effective potential diverges for $Z \to \pm \infty$.  If $\dot{Z}_{\rm int}^2 < V_{\rm crt}^2$, the effective potential is then positive and the resulting motion is periodic and confined within a finite spatial region with $0 <  \dot{Z}^2 \le \dot{Z}_{\rm int}^2$; moreover, the requirement of timelike motion, namely,  $1 - \dot{Z}^2 + k\,Z^2 > 0$ is satisfied. However, for $V_{\rm crt}^2 < \dot{Z}_{\rm int}^2 < 1$, the effective potential is negative and the motion accelerates until the timelike motion asymptotically approaches a null ray, in which case $1 - \dot{Z}^2 + k\,Z^2 = 0$. This speed of light singularity occurs at $Z = Z_{\rm N}$, where
\begin{equation}\label{I9}
 1+ 2\,k\,Z_{\rm N}^2 = \left(2\, \dot{Z}_{\rm int}^2 -1\right)\,e^{2\,k\,Z_{\rm N}^2}\,.
\end{equation}
The unique solution of this equation can be expressed by means of the Lambert $W$ function.  This function in the real domain has two branches $W_0(x)$ and $W_{-1}(x)$. The principal branch, $W_0$, is such that $W_0(x) \ge -1$, while $W_{-1}(x) \le -1$.  We find 
\begin{equation}\label{I10}
 1+ 2\,k\,Z_{\rm N}^2 = - W_{-1}(\gamma)\,, \qquad \gamma := \frac{1}{e}\, \left(1 - 2\, \dot{Z}_{\rm int}^2\right)\,, \qquad -\frac{1}{e} < \gamma < 0\,.
\end{equation}
In connection with tidal acceleration to the speed of light, we note that as the motion approaches the boundary of the Fermi coordinate patch, the higher-order terms that we have thus far neglected come into play and moderate the approach to the speed of light. This issue has been discussed in detail in Refs.~\cite{Chicone:2005da, Chicone:2005vn}. The mitigation of the speed of light singularity has been explicitly demonstrated using exact Fermi coordinates in special cases; however, only approximate analytic treatments are possible in the Schwarzschild and Kerr spacetimes~\cite{Chicone:2005vn, BGJ}. 

Let us next assume that $k<0$. In this case, the effective potential~\eqref{I8} exponentially goes to zero as $Z \to \pm \infty$.  For $\dot{Z}_{\rm int}^2 < V_{\rm crt}^2$, the motion accelerates 
toward $\dot{Z}^2 = V_{\rm crt}^2$; similarly, for $ \dot{Z}_{\rm int}^2 > V_{\rm crt}^2$, the motion decelerates toward $\dot{Z}^2 = V_{\rm crt}^2$. This means that for $k<0$, the critical solutions of the tidal equation, 
$Z = \pm V_{\rm crt}\, T$, are \emph{attractors}. That is, tidal motions with speeds above or below the critical speed eventually tend to approach the critical speed. Further discussion of this circumstance can be found in Ref.~\cite{Chicone:2002kb}.

We now turn to a detailed discussion of the analytic solutions of Eqs.~\eqref{A14} and~\eqref{A15} in the Kerr and Schwarzschild spacetimes. Greek indices run from $0$ to $3$, while Latin indices run from $1$ to $3$. The signature of the spacetime metric is $+2$.

\section{Exterior Kerr Spacetime: Observer at Rest on the Axis of Rotation}

We are interested in the family of test observers that are spatially at rest on the axis of rotational symmetry of the exterior Kerr spacetime. The Kerr metric is given in Boyer-Lindquist coordinates $(t,r,\theta,\phi)$ by~\cite{Chandra}
\begin{align}\label{Ka}
\nonumber -ds^2 = {}& -\left(1-\frac{2\,M\,r}{\Sigma}\right)\,dt^2 - 4\,\frac{M\,a\,r}{\Sigma} \sin^2\theta\,dt\,d\phi +\frac{\Sigma}{\Delta}\,dr^2+\Sigma\, d\theta^2 \\
&+\left(r^2+a^2 + \frac{2\,M\,a^2\,r}{\Sigma}\,\sin^2\theta \right)\,\sin^2\theta\, d\phi^2\,.
\end{align}
Here, $M$ and $a$ are the mass and specific angular momentum of the Kerr source, respectively. Moreover, $\Sigma=r^2+a^2\cos^2\theta$ and $\Delta=r^2-2Mr+a^2$. The spacetime is asymptotically flat and reference observers at rest in space exist from the asymptotic region all the way down to the exterior of the stationary limit surface ($\Sigma = 2\,M\,r$). 

 A fiducial observer that is at rest on the axis of symmetry carries a nonrotating orthonormal tetrad frame $\lambda^{\mu}{}_{\hat \alpha}$ such that 
\begin{equation}\label{Kb}
\lambda_{\hat 0}  = \left(\frac{r^2 + a^2}{\Delta}\right)^{1/2}\,\partial_t\,, \qquad  \lambda_{\hat 1}  = \left(\frac{\Delta}{r^2 + a^2}\right)^{1/2}\,\partial_r\,
\end{equation}
and $(\lambda_{\hat 2}, \lambda_{\hat 3})$ are unit vectors in the plane orthogonal to the rotation axis. The rotational symmetry of the Kerr field about its axis leads to enormous simplification of the general analysis presented in Appendices B and C. Specifically, we have $\lambda^{\mu}{}_{\hat 0} = \bar{u}^\mu = d\bar{x}^\mu/d\tau$, which is the 4-velocity of the fiducial observer and $\tau$ is its proper time. Moreover, the observer's acceleration 
$D\lambda^{\mu}{}_{\hat 0}/d\tau = \mathbf{A}^\mu = A^{\hat i}\,\lambda^{\mu}{}_{\hat i}$ is given by
\begin{equation}\label{K1}
(A_{\hat 1}, A_{\hat 2}, A_{\hat 3})  = (A, 0, 0)\,, \qquad  A = \frac{M(r^2-a^2)}{(r^2 + a^2)^{3/2}\,(r^2-2Mr+ a^2)^{1/2}}\,.
\end{equation}

We establish a Fermi normal coordinate system along the world line of the fiducial observer (see Appendix A).  The Fermi coordinates $X^{\hat \mu} = (T, X, Y, Z)$ in this case are such that $X$ denotes the radial coordinate along the Kerr symmetry axis, while $(Y, Z)$ are the two orthogonal transverse coordinates. 
In the tidal Eq.~\eqref{A14}, the relevant acceleration parameters are then given by Eq.~\eqref{K1}, while the gravitoelectric and the gravitomagnetic curvature components are given by $\mathcal{E}$ = diag$(-2E, E, E)$ and $\mathcal{H}$ = diag$(-2H, H, H)$, where
\begin{equation}\label{K2}
E := \mathbb{E}(\theta = 0) = \frac{Mr(r^2-3a^2)}{(r^2 + a^2)^3}\,, \qquad  H := \mathbb{H}(\theta = 0) = -\frac{Ma(3r^2-a^2)}{(r^2 + a^2)^3}\,;
\end{equation}
see Appendix C. 

After these preliminary considerations, we now turn to the tidal equations of motion~\eqref{A14} to linear order in $(X, Y, Z)$, namely,  
\begin{align}\label{K3}
\nonumber  \frac{d^2X}{dT^2} {}&+ (A -2E\,X)\,(1-2\dot{X}^2) +A^2\,X\,(1+2\dot{X}^2)  \\   
 &-\tfrac{2}{3}\,E\, [X(\dot{Y}^2 + \dot{Z}^2) + 2\,\dot{X}(Y\dot{Y} + Z\dot{Z})] + 2\,H (1-\dot{X}^2)(Y\,\dot{Z} - Z\,\dot{Y}) = 0\,,
\end{align}
\begin{align}\label{K4}
\nonumber  \frac{d^2Y}{dT^2} {}&+ E\,Y\,(1-2\dot{Y}^2) -2\,[A- (A^2 + \tfrac{7}{3}\,E)\,X]\,\dot{X}\,\dot{Y} -\tfrac{2}{3}\,E\, [Y(\dot{X}^2 - 2\,\dot{Z}^2) + 5\,Z\,\dot{Y}\,\dot{Z}] \\   
 & + 2\,H\, [2\,X\,\dot{Z} + Z\,\dot{X} - \dot{X}\,\dot{Y}(Y\,\dot{Z} - Z\,\dot{Y})] = 0\,,
\end{align}
\begin{align}\label{K5}
\nonumber  \frac{d^2Z}{dT^2} {}&+ E\,Z\,(1-2\dot{Z}^2) -2\,[A- (A^2 + \tfrac{7}{3}\,E)\,X]\,\dot{X}\,\dot{Z} -\tfrac{2}{3}\,E\, [Z(\dot{X}^2 - 2\,\dot{Y}^2) + 5\,Y\,\dot{Y}\,\dot{Z}] \\   
 & - 2\,H\, [2\,X\,\dot{Y} + Y\,\dot{X} + \dot{X}\,\dot{Z}(Y\,\dot{Z} - Z\,\dot{Y})] = 0\,.
\end{align}
To these equations we must add the condition that the tidal motion is timelike, namely, 
\begin{align}\label{K6}
\nonumber  {}& (1+ A\,X)^2 - (\dot{X}^2 + \dot{Y}^2+ \dot{Z}^2)- E\, (2\,X^2-Y^2-Z^2) \\
 & +\tfrac{1}{3}\,E\,[ 2\,(Y\,\dot{Z} - Z\,\dot{Y})^2 - (X\,\dot{Z} - Z\,\dot{X})^2 - (X\,\dot{Y} - Y\,\dot{X})^2] + 4\,H X\, (Y\,\dot{Z} - Z\,\dot{Y}) > 0\,.
\end{align}

The azimuthal symmetry of Kerr geometry is reflected in these approximate equations. That is, keeping $(X, \dot{X})$ unchanged at a given time $T$, one can show that the $(Y, Z)$ system of equations is invariant under a rotation by a constant angle about the axis of symmetry.  Moreover, inspection of the $(Y, Z)$ system reveals that $Y(T) = 0$ and $Z(T) = 0$ are not separately possible solutions so long as $H \ne 0$. We note from Eq.~\eqref{K2} that $H = 0$ if either $a = 0$, so that the source is spherically symmetric and described by the Schwarzschild spacetime, or $r = a /\sqrt{3}$. In the latter case, the observer must be fixed on the rotation axis at $r = a /\sqrt{3}$, which is possible in the exterior Kerr spacetime if $2\,a > \sqrt{3}\,M$.

\subsection{Motion Along the Rotation Axis of Kerr Spacetime}

If $Y(T) = Z(T) = 0$, then our system reduces to 
\begin{equation}\label{K7}
 \frac{d^2X}{dT^2}+ (A -2E\,X)\,(1-2\dot{X}^2) +A^2\,X\,(1+2\dot{X}^2) = 0\,
\end{equation}
and
\begin{equation}\label{K8}
(1+ A\,X)^2 - \dot{X}^2 - 2\,E\,X^2 > 0\,,
\end{equation}
which describe tidal motion along the Kerr rotation axis. In the absence of acceleration ($A = 0$), Eq.~\eqref{K7} reduces to Eq.~\eqref{I1} for the case of constant $k = -2\,E < 0$, where the critical currents turn out to be attractors as discussed in Section II. However, the presence of acceleration drastically changes the dynamical behavior of the system. To see how this comes about in some detail, let us define the Fermi velocity of a free test particle $V$, 
\begin{equation}\label{K9}
V := \frac{dX}{dT} = \dot{X}\,
\end{equation}
and write Eq.~\eqref{K7} in the form
\begin{equation}\label{K10}
 \frac{dV^2}{dX}+ 4\,[(2\,E + A^2)\,X - A ] \,V^2 + 2\,[A - (2\,E - A^2)\,X] = 0\,.
\end{equation}
It is evident that our $(X, V)$ system has a \emph{rest point} given by $(X_{\rm rp}, 0)$, where
\begin{equation}\label{K11}
X_{\rm rp} = \frac{A}{2\,E - A^2}\,.
\end{equation}
As expected, for $A = 0$ we recover $(X, V) = (0, 0)$, which is the rest point associated with Eq.~\eqref{I4}.

It is straightforward to integrate Eq.~\eqref{K10} by means of an integrating factor $\mathcal{I}$, 
\begin{equation}\label{K12}
\mathcal{I}(X) = \exp{[2\,(2\,E + A^2)\,X^2 - 4\,A\,X]}\,, \qquad \frac{d\mathcal{I}}{dX} = 4\,[(2\,E + A^2)\,X - A]\,\mathcal{I}\,. 
\end{equation}
With the initial condition that at $T = 0$  a free test particle at $X = 0$ has Fermi coordinate velocity  $V = V_0$, $|V_0| < 1$, Eq.~\eqref{K10} can be integrated once and we find the ``energy" equation
\begin{equation}\label{K13}
\mathcal{I}(X) \,V^2 +\mathcal{W}_{\rm eff}(X) = V_0^2\,, 
\end{equation}
where $\mathcal{W}_{\rm eff}(X)$ is the effective potential energy in this case given by
\begin{equation}\label{K14}
\mathcal{W}_{\rm eff}(X) =2\,\int_0^X [A - (2\,E - A^2)\,\xi]\,\mathcal{I}(\xi) \,d\xi\,.
\end{equation}
The path of the free particle $X(T)$ can be simply obtained from integrating Eq.~\eqref{K13}. To clarify the nature of this motion, we note that $\mathcal{I}(X) > 0$; hence, Eq.~\eqref{K13} implies that the motion is confined to the region 
\begin{equation}\label{K15}
\mathcal{W}_{\rm{eff}}(X)\le V_0^2\,.
\end{equation}
It follows from the definition of the effective potential~\eqref{K14} that this function vanishes at $X = 0$ and has an extremum at $X = X_{\rm rp}$. To go further, we need to specify the physical parameters of the configuration under discussion here. 

\subsection{$r \gg M$ and $r \gg a$}

In most astrophysical situations of interest, we deal with motions of energetic particles relative to an environment that constitutes the rest frame of the gravitational source. 
Employing standard techniques, motions are usually observable that take place at a distance $r$ from the source such that $r \gg M$ and $r \gg a$. We therefore focus our attention on such situations and note that 
\begin{equation}\label{K16}
A \approx \frac{M}{r^2}\, \left(1 + \frac{M}{r} - 3\,\frac{a^2}{r^2} + \cdots   \right)\,, \qquad E \approx \frac{M}{r^3}\, \left(1 - 6\,\frac{a^2}{r^2} + \cdots   \right)\,,
\end{equation}
where we have neglected higher-order terms in the small quantities $M/r$ and $a^2/r^2$. Similarly, we can write
\begin{equation}\label{K17}
X_{\rm rp} \approx \frac{r}{2}\, \left(1 + \frac{3}{2}\,\frac{M}{r} + 3\,\frac{a^2}{r^2} + \cdots   \right)\,.
\end{equation}

It follows from these results and inspection of the expression for the effective potential energy function $\mathcal{W}_{\rm eff}(X)$ that we have a simple potential barrier here that starts from zero at $X = 0$ and  reaches its maximum value of  $\nu_{\rm crt}^2$ at $X_{\rm rp}$, where $\nu_{\rm crt} > 0$ and
\begin{equation}\label{K18}
\nu_{\rm crt}^2 := \mathcal{W}_{\rm{eff}}(X_{\rm rp}) =2\,\int_0^{X_{\rm rp}} [A - (2\,E - A^2)\,\xi]\,\mathcal{I}(\xi) \,d\xi\,.
\end{equation}
The energy Eq.~\eqref{K13} implies that at the rest point $(X, V) = (X_{\rm rp}, 0)$, $V_0^2 = \nu_{\rm crt}^2 $.

Introducing $\xi = (r/2) \mu$ in the integrand of Eq.~\eqref{K18} and using only the dominant terms in $A$ and $E$, we find
\begin{equation}\label{K19}
\nu_{\rm crt}^2 \approx \frac{M}{r}\, \int_{0}^{1} (1-\mu)\,e^{\frac{M}{r}\,(\mu^2-2\,\mu)} \,d\mu\,.
\end{equation}
Therefore,  to lowest order in the small quantities, we have
\begin{equation}\label{K20}
\nu_{\rm crt}^2 \approx \frac{M}{2\,r}\,.
\end{equation}
With these results for the effective potential, we now return to the analysis of the energy Eq.~\eqref{K13}.

\subsection{Tidal Acceleration}

Imagine a free test particle at the position of the fiducial observer ($X = 0$) with initial velocity $V_0$,  $1 > V_0 > 0$,  such that $V_0^2 < \nu_{\rm crt}^2$; that is, in Eq.~\eqref{K13}, $V_0^2$ is less than the maximum height of the potential barrier. In this case, the particle reaches a certain maximum distance $X_{\rm max}$, $ 0 < X_{\rm max} < X_{\rm rp}$,  along the radial direction, has a turning point at 
$V_0^2 =  \mathcal{W}_{\rm{eff}}(X_{\rm max})$ and falls back toward the gravitational source. On the other hand, for  $V_0^2 > \nu_{\rm crt}^2$, the particle tidally accelerates toward the speed of light. Of course, near the boundary of the Fermi coordinate patch, higher-order tidal terms intervene and mitigate the speed of light singularity, which would occur when $V^2 = (1+AX)^2 -2E\, X^2$ in accordance with Eq.~\eqref{K8}.  Nevertheless, considerable tidal acceleration is expected to take place in practice. The implications of this result for the tidal acceleration of astrophysical jets have been extensively studied in Ref.~\cite{Bini:2017uax}, where various possibilities have been thoroughly investigated.  Indeed, Eqs.~\eqref{K7} and~\eqref{K8} are invariant under the transformation $(X, A, E) \mapsto (-X, -A, E)$ in conformity with the double jet structure of astrophysical outflows. The present mechanism for the tidal acceleration of outflows is independent of the presence or absence of an event horizon. As emphasized in Ref.~\cite{Bini:2017uax}, the comparison of these theoretical results with observational data must include electromagnetic forces as well~\cite{BPu}.  

It is instructive to restate our main physical results here in Newtonian terms. The reference observer is fixed on the rotation axis at the Boyer-Lindquist radial coordinate $r$ sufficiently far from the Kerr source such that  $r \gg M$ and $r \gg a$. The Newtonian escape velocity at the position of reference observer is
\begin{equation}\label{K21}
V_{\rm esc} := \left(\frac{2\,G\,M}{r}\right)^{1/2}\,.
\end{equation}
It follows from Eq.~\eqref{K20} that $\nu_{\rm crt}$ is approximately equal to one-half of the Newtonian escape velocity; that is, 
\begin{equation}\label{K22}
\nu_{\rm crt} \approx \frac{1}{2}\, V_{\rm esc}\,.
\end{equation}   
Thus if a free test particle is launched outward from the position of the reference observer with a velocity less than about one-half of the Newtonian escape velocity, then the test particle cannot escape the gravitational field of the Kerr source; in fact, as expected, it reaches a maximum height and then falls back toward the source. However, if the initial velocity is more than about one-half of the Newtonian escape velocity, then the particle tidally accelerates away from the Kerr source. 

Let us briefly digress here and mention that for purely radial motion in the exterior Schwarzschild spacetime, the escape velocity for a free test particle starting at Schwarzschild radial coordinate $r$ is indeed given by Eq.~\eqref{K21}, where the escape speed of the particle is measured by the fiducial observer that is fixed at $r$. On the other hand,  our treatment in this section employs the invariantly defined quasi-inertial Fermi normal coordinate system. 

We have thus far considered motion along the axis of symmetry of a Kerr source. For $a = 0$, the Kerr metric reduces to the spherically symmetric Schwarzschild metric, in which case our treatment applies to motion along any radial direction. The corresponding expressions for $A$ and $E$ simplify for $a = 0$ in Eqs.~\eqref{K1} and~\eqref{K2}, respectively, but our main results remain unchanged. Indeed, sufficiently far from any astronomical  source, we expect that its gravitational field is dominated by its mass $M$ and the accelerated outflow can occur above the threshold~\eqref{K22} along any radial direction away from mass $M$. 

In our discussion of the solutions of the tidal equations, azimuthal symmetry of the exterior Kerr spacetime has made it possible to do analytic work along the axis of rotational symmetry. This treatment can be extended to any radial direction in the exterior Schwarzschild spacetime due to its spherical symmetry.   This symmetry, furthermore, makes it possible to go beyond the radial direction $X$ in the next section and consider the solutions of the tidal equations in the $(X, Y)$ plane with no loss in generality.

\section{Exterior Schwarzschild Spacetime}

In this section, we assume $a = 0$, so that the reference observer is at rest in the spherically symmetric exterior of Schwarzschild spacetime. Following the discussion of the previous section, we can now set $Z(T) = 0$ and consider motion in the $(X, Y)$ plane. The equations of motion then reduce to 
\begin{align}\label{S1}
\nonumber  \frac{d^2X}{dT^2} {}&+ (A -2E\,X)\,(1-2\dot{X}^2) +A^2\,X\,(1+2\dot{X}^2)  \\   
 &-\tfrac{2}{3}\,E\,\dot{Y}\,(X\,\dot{Y} + 2\, Y\,\dot{X}) = 0\,
\end{align}
and
\begin{align}\label{S2}
\nonumber  \frac{d^2Y}{dT^2} {}&+ E\,Y\,(1-\tfrac{2}{3}\,\dot{X}^2-2\,\dot{Y}^2) \\
&-2\,[A- (A^2 + \tfrac{7}{3}\,E)\,X]\,\dot{X}\,\dot{Y}= 0\,,
\end{align}
while the timelike condition reduces to
\begin{align}\label{S3}
\nonumber  {}& (1+ A\,X)^2 - (\dot{X}^2 + \dot{Y}^2)- E\, (2\,X^2-Y^2) \\
 & -\tfrac{1}{3}\,E\,(X\,\dot{Y} - Y\,\dot{X})^2 > 0\,.
\end{align}

Inspection of Eqs.~\eqref{S1} and~\eqref{S2} reveals that an analytic solution of this system is possible and is given by a \emph{rest point} $(X, \dot{X}) = (X_{\rm S}, 0)$ along the radial direction and a \emph{critical current} along the transverse direction. That is,  
\begin{equation}\label{S4}
 X_{\rm S} = \frac{3\,A}{7\,E-3A^2} = 3\,\frac{[r^3(r-2M)]^{1/2}}{7\,r-17\,M}\,
\end{equation}
for $r \ne 17M/7$ and 
\begin{equation}\label{S5}
Y = Y_0 \pm V_{\rm crt}\,(T-T_0)\,,
\end{equation}
where $Y_0$ and $T_0$ are constants.  For $r \gg M$, we have 
\begin{equation}\label{S5a}
X_{\rm S} = \frac{3}{7}\,r \,\left(1+ \frac{10}{7}\,\frac{M}{r} + \cdots \right)\,,
\end{equation}
and the timelike condition~\eqref{S3} is satisfied in this case. This \emph{exact} solution of the $(X,Y)$ system means that the Fermi coordinate along the radial direction $X$ remains constant while a critical tidal current $Y$ flows in the transverse direction. 

It has not been possible  to find other exact solutions of the $(X, Y)$ system. To proceed further, we perturb the $(X, Y)$ system to linear order about the exact solution given in Eqs.~\eqref{S4} and~\eqref{S5} and then solve the resulting linear perturbation equations via expansions in power series.

Let us define a new temporal variable $\eta$,
\begin{equation}\label{S6a}
 \eta := Y_0 \pm V_{\rm crt}\,(T-T_0)\,
\end{equation}
and substitute 
\begin{equation}\label{S6}
X(\eta) = X_{\rm S} + \epsilon\,P(\eta)\,, \qquad Y(\eta) = \eta +  \epsilon\,Q(\eta)\,\,,
\end{equation}
in Eqs.~\eqref{S1} and~\eqref{S2}. Here $\epsilon$ , $0 < \epsilon \ll 1$ is a constant perturbation parameter. To first order in $\epsilon$, we find
\begin{equation}\label{S7}
\frac{d^2P}{d\eta^2} -\tfrac{4}{3} E \eta\, \frac{dP}{d\eta} - 2 \tfrac{A}{X_{\rm S}}\, P - \tfrac{4}{3} E X_{\rm S}\, \frac{dQ}{d\eta} = 0\,,
\end{equation}
\begin{equation}\label{S8}
\frac{d^2Q}{d\eta^2} - 4 E \eta\, \frac{dQ}{d\eta} +4\, A^2 X_{\rm S}\, \frac{dP}{d\eta} = 0\,.
\end{equation}
Substituting Eq.~\eqref{S7} and its derivative in Eq.~\eqref{S8}, we find the third-order homogeneous ordinary differential equation
\begin{equation}\label{S9}
\frac{d^3P}{d\eta^3} - w_1 \eta\, \frac{d^2P}{d\eta^2} + (w_2\,\eta^2 - w_3)\,\frac{dP}{d\eta} +w_4 \, \eta\,P = 0\,,
\end{equation}
where $w_1, w_2, w_3$ and $w_4$ are \emph{constant} coefficients that are given by
\begin{equation}\label{S10}
w_1 = \frac{16}{3}\,E\,, \quad w_2 = \frac{16}{3}\,E^2\,, \quad w_3 = 2\,E\, \left[\frac{3r-7M}{r-2M} -24\,\frac{M^2}{(7r-17M)^2}\right]\,
\end{equation}
and
\begin{equation}\label{S11}
 w_4 =\frac{8}{3}\,E^2\,\left(\frac{7r-17M}{r-2M}\right)\,, \qquad E = \frac{M}{r^3}\,.
\end{equation}
The solutions of Eq.~\eqref{S9} form a linear manifold. Indeed, the general solution of this homogeneous linear differential equation can be expressed as a linear sum with constant coefficients of three independent solutions~\cite{CH}. 

To find explicit solutions of Eq.~\eqref{S9}, we resort to infinite series in  powers of $\eta$.  As in the method of Frobenius, we assume
\begin{equation}\label{S12}
P(\eta) = \eta^{\rho}\, (a_0 + a_1\, \eta + a_2\, \eta^2 + a_3\, \eta^3 + \cdots)\,,
\end{equation}
where $a_n$, $n = 0, 1, 2, \cdots$, are constants and $a_0 \ne 0$ by definition. Substitution of this series in Eq.~\eqref{S9} results in an infinite series in increasing powers of $\eta$ starting with $\eta^{\rho - 3}$. As in the standard procedure, we recover the relations
\begin{equation}\label{S13}
\rho \,(\rho - 1)\, (\rho - 2)\,a_0 = 0\,,
\end{equation}
\begin{equation}\label{S14}
\rho \,(\rho^2 - 1)\,a_1 = 0\,,
\end{equation}
\begin{equation}\label{S15}
\rho \,(\rho + 1)\, (\rho + 2)\,a_2 = \rho \,[(\rho - 1)\,w_1 +  w_3]\,a_0\,,
\end{equation}
\begin{equation}\label{S16}
(\rho + 1) \,(\rho + 2)\, (\rho + 3)\,a_3 = (\rho + 1)\,(\rho\,w_1 + w_3)\,a_1\,
\end{equation}
and the recurrence relation
\begin{align}\label{S17}
\nonumber (n + \rho + 2) \,{}&(n + \rho + 3)\, (n + \rho + 4)\,a_{n + 4} = {}&   \\
{}&(n + \rho + 2) \,[(n + \rho + 1)\,w_1 + w_3]\,a_{n + 2} - [(n + \rho) \, w_2 + w_4]\,a_n\,
\end{align}
for $n = 0, 1, 2, \cdots$. 

It follows from Eq.~\eqref{S13} that $\rho = 0, 1$ or 2, since $a_0 \ne 0$. In each case, we can find power series solutions for $P(\eta)$ that can be expressed  as superpositions of solutions $\mathbb{S}_0$, $\mathbb{S}_1$ and $\mathbb{S}_2$ defined by
\begin{equation}\label{S18}
\mathbb{S}_0(\eta) :=  1 - \frac{w_4}{4!}\,\eta^4 - \frac{(3\,w_1 + w_3)\,w_4}{6!}\,\eta^6 + \cdots\,,
\end{equation}
\begin{equation}\label{S19}
\mathbb{S}_1(\eta) := \eta + \frac{1}{3!} w_3\, \eta^3 +  \frac{1}{5!} w_5\, \eta^5 + \cdots \, 
\end{equation}
and 
\begin{equation}\label{S20} 
\frac{1}{2}\,\mathbb{S}_2(\eta) := \frac{1}{2!}\,\eta^2+  \frac{1}{4!} \,(w_1 +w_3)\, \eta^4 + \frac{1}{6!}\,w_6\,\eta^6 +\cdots\,,
\end{equation}
where $w_5$ and $w_6$ are  constants  defined by
\begin{equation}\label{S21}
w_5 := 2\,w_1\,w_3 - 2\,w_2 + w_3^2 -2\,w_4\,
\end{equation}
and
\begin{equation}\label{S22}
w_6 := (3\,w_1 + w_3)(w_1 + w_3) - 3 (2\,w_2 + w_4)\,.
\end{equation}
For $\rho = 0$, it turns out that $P(\eta)$ can be written as a linear superposition of the three independent solutions $\mathbb{S}_0$, $\mathbb{S}_1$ and $\mathbb{S}_2$ with constant coefficients $a_0$, $a_1$ and $a_2$, respectively. On the other hand, for $\rho = 1$, $P(\eta)$ is a linear superposition of $\mathbb{S}_1$ and $\mathbb{S}_2$ with constant coefficients $a_0$ and $a_1$, respectively, while for $\rho = 2$, $P(\eta)$ is simply proportional to $\mathbb{S}_2$ with proportionality constant equal to $a_0$.

Given a solution $P(\eta)$ of Eq.~\eqref{S9}, we can use Eq.~\eqref{S7} to find a corresponding solution for $Q(\eta)$. To illustrate the general case, we assume 
\begin{equation}\label{S23}
P(\eta) = a_0\,\mathbb{S}_0(\eta) + a_1\, \mathbb{S}_1(\eta) + a_2\, \mathbb{S}_2(\eta)\,,
\end{equation}
corresponding to  $\rho = 0$. 
From Eq.~\eqref{S7}, we can work out the infinite series solution corresponding to $Q(\eta)$, namely, 
\begin{align}\label{S24}
\nonumber Q(\eta) = {}&Q_0  + \frac{a_0}{q_1}\,\left(\frac{w_4}{8\,E}\right)\,\left(\eta + \frac{2}{3}\,E\,\eta^3 + \cdots \right)     \\
{}& - \frac{a_1}{q_1}\,\left(\frac{1}{2!}\,q_2\,\eta^2 + \frac{1}{4!}\,q_4\,\eta^4 + \cdots \right) - \frac{a_2}{q_1}\,\left(\eta +\frac{1}{3!}\,q_3\,\eta^3 + \cdots \right)\,,
\end{align}
where $Q_0$ is an integration constant and 
\begin{equation}\label{S25}
q_1 = - 2\,E\,\frac{[r^3(r-2M)]^{1/2}}{7\,r-17\,M}\,, \qquad q_2 = \frac{1}{2}\,w_3 -  E\,\left(\frac{3r-7M}{r-2M}\right)\,, 
\end{equation}
\begin{equation}\label{S26}
q_3 = w_1 + w_3 - \frac{2}{3}\,E\,\left(\frac{11r-25M}{r-2M}\right)\,, \qquad  q_4 = \frac{1}{2}\,w_5 - \frac{1}{3}\,w_3 \,E\,\left(\frac{13r-29M}{r-2M}\right)\,.
\end{equation}

Using expressions~\eqref{S23} and~\eqref{S24} for $P(\eta)$ and $Q(\eta)$, respectively,  in Eq.~\eqref{S6},  we have in the neighborhood of the critical solution an approximate power series solution of the tidal equations in the $(X, Y)$ plane that depends on four independent parameters, namely, $\epsilon\,a_0$, $\epsilon\,a_1$, $\epsilon\,a_2$ and $\epsilon\,Q_0$. We note that a general solution in terms of the temporal parameter $\eta$ is expected to depend on four independent parameters corresponding to the initial position and velocity in the $(X, Y)$ plane. 
The issue of the convergence of such power series solutions as well as their physical interpretations requires further investigation that is beyond the scope of this work.

\section{Discussion}

In GR, the generalized Jacobi equation is the exact analogue of the Lorentz force law of electrodynamics. Just as the electric and magnetic fields can be measured via the motion of charged particles using the Lorentz force law, the components of the Riemann curvature tensor can be measured, in principle, using the generalized Jacobi equation~\cite{Mashhoon:2018lgk}. This equation contains a critical speed $V_{\rm crt} = c/\sqrt{2} \approx 0.71 \,c$  and the associated critical currents that are explored in the present paper. In particular, we have studied the generalized Jacobi equation for the motion of free test particles with respect to the rest frame of a central Kerr source. Among other things, we have elucidated a general relativistic tidal acceleration mechanism that is relevant for the theory of astrophysical jets~\cite{Bini:2017uax}. Moreover, the approach adopted in this work can be employed to determine the gravitational influence of the quiescent massive black hole at the Galactic Center on the dynamics of the energetic particles involved in the production of giant bipolar radio bubbles near the Galactic Center~\cite{FYZ}.

\appendix

\section{Jacobi-Type Equation in Fermi Coordinates}

The purpose of this appendix is to consider a gravitational field and the associated curved spacetime manifold within the GR framework.  An arbitrary test accelerated observer in this spacetime  carries along its path a nonrotating orthonormal tetrad frame. We establish a Fermi normal coordinate system along the world line of this reference observer and investigate the motion of free test particles in the Fermi system. We derive and study the resulting Jacobi-type equation. 

\subsection{Fermi Coordinates}

We imagine an accelerated test observer with proper time $\tau$ following a timelike path $\bar{x}^\mu(\tau)$ in spacetime. The reference observer carries an orthonormal Fermi-Walker transported tetrad frame $\lambda^{\mu}{}_{\hat \alpha} (\tau)$ along its world line. Here, $\lambda^{\mu}{}_{\hat 0} (\tau) = d\bar{x}^\mu(\tau)/d\tau =  \bar{u}^\mu(\tau)$ is the unit 4-velocity vector of the observer, 
$D\bar{u}^\mu(\tau)/d\tau =  \mathbf{A}^{\mu}$ is its acceleration  and $\lambda^{\mu}{}_{\hat i} (\tau)$, $i = 1, 2, 3$, constitute the local spatial frame of the observer.  At any given event $\tau$ along this path, we consider all spacelike geodesics that originate from this event and are orthogonal to the observer's world line. These form a \emph{local} spacelike hypersurface. Let $x^\mu$ be an event on this hypersurface such that there is a \emph{unique} spacelike geodesic of proper length $\sigma$ that connects $x^\mu$ to the observer's world line at $\bar{x}^\mu(\tau)$. If $\xi^\mu(\tau)$ is the unit spacelike vector tangent to this geodesic at event $\tau$ along the world line, then $\xi^\mu(\tau)\,\bar{u}_{\mu}(\tau) = 0$.  We assign Fermi normal coordinates $X^{\hat \mu} = (T, X^{\hat i})$ to event $x^\mu$, where
\begin{equation} \label{A1}
X^{\hat 0} = T := \tau\,, \qquad X^{\hat i} := \sigma\, \xi^\mu(\tau)\, \lambda_{\mu}{}^{\hat i}(\tau)\,.
\end{equation}

The fiducial observer has $X^{\hat i} = 0$ and is thus permanently fixed at the spatial origin of the Fermi coordinate system. Let us note that for $i = 1, 2, 3$, $\xi^\mu(\tau)\, \lambda_{\mu}{}^{\hat i}(\tau)$ constitute the direction cosines at proper time $\tau$; therefore,  Fermi coordinates are the natural extension of the inertial Cartesian coordinates to the curved spacetime of GR~\cite{Synge}. Fermi coordinates are admissible in a finite cylindrical domain about the world line of the fiducial observer with $|\mathbf{X}| \ll L(T)$, where $L(T)$ is a certain  infimum of the observer's acceleration length and the radius of curvature of spacetime.

The spacetime metric in Fermi coordinates is given by $-ds^2= g_{\hat \mu \hat \nu}\,dX^{\hat \mu} \,dX^{\hat \nu}$, where~\cite{Mash77}
\begin{equation}\label{A2}
g_{\hat 0 \hat 0} = -1 -2A_{\hat i}\,X^{\hat i} - (A_{\hat i}\,A_{\hat j} + R_{\hat 0 \hat i \hat 0 \hat j})\,X^{\hat i}\,X^{\hat j} + O(|\mathbf{X}|^3)\,,
\end{equation}
\begin{equation}\label{A3}
g_{\hat 0 \hat i} = -\frac{2}{3} \, R_{\hat 0 \hat j \hat i \hat k}\,X^{\hat j}\,X^{\hat k} + O(|\mathbf{X}|^3)\,,
\end{equation}
\begin{equation}\label{A4}
g_{\hat i \hat j} = \delta_{\hat i \hat j} -\frac{1}{3} \,R_{\hat i \hat k \hat j \hat l}\,X^{\hat k}\,X^{\hat l} + O(|\mathbf{X}|^3)\,.
\end{equation}
Here, we have defined 
\begin{equation}\label{A5}
A_{\hat i} (T)  := \mathbf{A}_{\mu}\,\lambda^{\mu}{}_{\hat i}\,
\end{equation}
and
\begin{equation}\label{A6}
R_{\hat \alpha \hat \beta \hat \gamma \hat \delta}(T) := R_{\mu \nu \rho \sigma}\,\lambda^{\mu}{}_{\hat \alpha}\,
\lambda^{\nu}{}_{\hat \beta}\,\lambda^{\rho}{}_{\hat \gamma}\,\lambda^{\sigma}{}_{\hat \delta}
\end{equation}
to be the components of the acceleration and the Riemann curvature tensor as measured by the reference observer, respectively. 

The Christoffel symbols in Fermi coordinates can be computed from the metric; indeed, the nonzero components to linear order in $|\mathbf{X}|$ can be obtained from~\cite{Mash77}
\begin{equation}\label{A7}
\Gamma^{\hat 0}_{\hat 0 \hat 0} = \frac{dA_{\hat i}}{d\,T}\,X^{\hat i}\,,\qquad \Gamma^{\hat 0}_{\hat 0 \hat i} = A_{\hat i}+(R_{\hat 0 \hat i \hat 0 \hat j}-A_{\hat i}\,A_{\hat j})\,X^{\hat j}\,,
\end{equation}
\begin{equation}\label{A8}
\Gamma^{\hat 0}_{\hat i \hat j} = \frac{2}{3}\, R_{\hat 0(\hat i \hat j)\hat k}\,X^{\hat k}\,,\qquad \Gamma^{\hat i}_{\hat 0 \hat 0} =  A_{\hat i}+(R_{\hat 0 \hat i \hat 0 \hat j}+A_{\hat i}\,A_{\hat j})\,X^{\hat j}\,,
\end{equation}
\begin{equation}\label{A9}
\Gamma^{\hat i}_{\hat 0 \hat j} = -R_{\hat 0 \hat k \hat i \hat j}\,X^{\hat k}\,,\qquad \Gamma^{\hat i}_{\hat j \hat k} = -\frac{2}{3}\, R_{\hat i(\hat j \hat k)\hat l}\,X^{\hat l}\,.
\end{equation}

\subsection{Jacobi-Type Equation}

The geodesic equation of motion in Fermi coordinates takes the form
\begin{equation}\label{A10}
\frac{dU^{\hat \mu}}{ds} + \Gamma^{\hat \mu}_{\hat \alpha \hat \beta}\, U^{\hat \alpha}\,U^{\hat \beta} = 0\,,
\end{equation}
where the 4-velocity of the free test particle in the Fermi system can be written as
\begin{equation}\label{A11}
U^{\hat \mu} = \frac{dX^{\hat \mu}}{ds} = \Gamma (1, \mathbf{V})\,, \qquad \mathbf{V} = \frac{d\mathbf{X}}{dT}\,
\end{equation}
and the Lorentz factor $\Gamma = dT/ds$ can be determined via $U^{\hat \mu}\,U_{\hat \mu} = -1$. We find
\begin{equation}\label{A12}
\frac{1}{\Gamma^2} = - g_{\hat 0 \hat 0} - 2\,g_{\hat 0 \hat i}\, V^{\hat i} - g_{\hat i \hat j}\, V^{\hat i} \,V^{\hat j} > 0\,.
\end{equation}
Here, the Fermi velocity $\mathbf{V}$ is a \emph{coordinate velocity}; however, at the location of the reference observer $\mathbf{X} = 0$, we must have $|\mathbf{V}| <1$. 

It is now straightforward to derive the reduced geodesic equation~\cite{Chicone:2002kb}
\begin{equation}\label{A13}
\frac{d^2 X^{\hat i}}{dT^2}+\left(\Gamma^{\hat i}_{\hat \alpha \hat \beta}-\Gamma^{\hat 0}_{\hat \alpha \hat \beta}V^{\hat i} \right) \frac{dX^{\hat \alpha}}{dT}\frac{dX^{\hat \beta}}{dT}=0\,,
\end{equation} 
or, more explicitly,
\begin{eqnarray}\label{A14}
&&\frac{d^2X^{\hat i}}{dT^2}+A_{\hat i}+(R_{\hat 0 \hat i \hat 0 \hat l}+A_{\hat i}\,A_{\hat l})X^{\hat l}-\frac{dA_{\hat l}}{dT}\,X^{\hat l} V^{\hat i} \nonumber\\
&& \qquad\,{}{}-2\,R_{\hat 0 \hat l \hat i \hat j}V^{\hat j}X^{\hat l}-2\,[A_{\hat j}+(R_{\hat 0 \hat j \hat 0 \hat l}-A_{\hat j}A_{\hat l})\,X^{\hat l}]V^{\hat i}V^{\hat j} \nonumber\\
&& \qquad\,{}{} -\frac{2}{3} \left(R_{\hat i \hat j \hat k \hat l}+R_{\hat 0 \hat j \hat k \hat l}V^{\hat i} \right) X^{\hat l}V^{\hat j}V^{\hat k} + O(|\mathbf{X}|^2) = 0\,.
\end{eqnarray} 
The Lorentz $\Gamma$ factor now takes the form
\begin{align}\label{A15}
\frac{1}{\Gamma^2} = {}&  (1+A_{\hat i}\,X^{\hat i})^2 - \delta_{\hat i \hat j}\,V^{\hat i}\,V^{\hat j} + R_{\hat 0 \hat i \hat 0 \hat j}\,X^{\hat i}\,X^{\hat j}  \\  \nonumber
 & +\frac{4}{3}\, R_{\hat 0 \hat j \hat i \hat k}V^{\hat i}  X^{\hat j}\,X^{\hat k}+ \frac{1}{3}\,R_{\hat i \hat k \hat j \hat l}\,V^{\hat i}\,V^{\hat j}\,X^{\hat k}\,X^{\hat l} +O(|\mathbf{X}|^3)> 0\,.
\end{align}

It is clear from this treatment that simple generalizations are possible; for instance, we can consider accelerated motion of test particles in the Fermi coordinate system as well. 

\section{Observers at Rest in Exterior Kerr Spacetime}

We are interested in the family of test observers that are spatially at rest in the exterior Kerr spacetime. In our considerations, these fiducial observers represent the rest frame of the Kerr source, which is specified by a mass $M$ that rotates uniformly with angular momentum $J$. In Boyer-Lindquist coordinates $(t,r,\theta,\phi)$, the Kerr metric  can be expressed as~\cite{Chandra}
\begin{eqnarray}
\label{B1}
-ds^2&=&g_{\alpha\beta}\,dx^\alpha dx^\beta\nonumber\\
&=&-dt^2+\frac{\Sigma}{\Delta}\,dr^2+\Sigma\, d\theta^2 +(r^2+a^2)\sin^2\theta\, d\phi^2\nonumber\\
&& +\frac{2Mr}{\Sigma}\,(dt-a\sin^2\theta\, d\phi)^2\,,
\end{eqnarray}
where $a = J/(Mc)$ is the specific angular momentum parameter and
\beq\label{B2}
\Sigma=r^2+a^2\cos^2\theta\,,\qquad \Delta=r^2-2Mr+a^2\,.
\eeq
We emphasize that the tidal currents under scrutiny in this work are independent of the existence of an event horizon. For  $a\le M$, for instance,  the source is a Kerr \emph{black hole}, but this circumstance has no essential impact on the existence of the critical tidal currents in this case. 

Let $\bar{u}^\mu = d\bar{x}^\mu / d\tau$ be the unit timelike 4-velocity vector of a member of the family of observers at rest, where $\tau$ is the proper time along the world line. We therefore have
\begin{equation}\label{B3}
\bar{u}=\left(\frac{\Sigma}{\Sigma - 2\,M\,r}\right)^{1/2}\,  \partial_t\,, \qquad \tau =\left(\frac{\Sigma- 2\,M\,r}{\Sigma}\right)^{1/2}\,t\,,
\end{equation}
where we have assumed that $\tau=0$ at $t=0$. Only positive square roots are considered throughout. The reference observers exist outside the \emph{stationary limit surface} of Kerr spacetime given by $\Sigma - 2\,M\,r = 0$.

Next, we need to set up a nonrotating tetrad frame $\lambda^{\mu}{}_{\hat \alpha} (\tau)$ along the world line of our reference observer such that $\lambda^{\mu}{}_{\hat 0} (\tau) = \bar{u}^\mu(\tau)$. We do this in several steps. First, we set up an orthonormal tetrad frame $e^{\mu}{}_{\hat \alpha}$  at each event in the exterior Kerr spaetime such that 
\beq\label{B4}
e_{\hat 0}=\frac{1}{\sqrt{-g_{tt}}}\partial_t\,,  \quad e_{\hat 1}=\frac{1}{\sqrt{g_{rr}}} \partial_r\,, \quad  e_{\hat 2}=\frac{1}{\sqrt{g_{\theta\theta}}} \partial_\theta\,,
\quad e_{\hat 3}=\frac{1}{\left(g_{\phi\phi}-\frac{g_{t\phi}^2}{g_{tt}}\right)^{1/2}}\left(-\frac{g_{t\phi}}{g_{tt}} \partial_t+\partial_\phi\right)\,,
\eeq
where the tetrad axes are primarily along the  Boyer-Lindquist coordinate directions. From
\beq\label{B5}
-\frac{g_{t\phi}}{g_{tt}} = -2\,\frac{Mar}{\Sigma -2Mr}\, \sin^2\theta\,, \quad  \left(g_{\phi\phi}-\frac{g_{t\phi}^2}{g_{tt}}\right)^{1/2}= \left(\frac{\Sigma \Delta}{\Sigma -2Mr}\right)^{1/2}\,\sin \theta\,, \quad \sqrt{-g}= \Sigma\,\sin \theta\,,
\eeq
 we find
\begin{eqnarray}\label{B6}
\nonumber e_{\hat 0} &=& \left(\frac{\Sigma}{\Sigma- 2\,M\,r}\right)^{1/2}\,\partial_t\,, \qquad e_{\hat 1}= \left(\frac{\Delta}{\Sigma}\right)^{1/2}\,\partial_r\,, \qquad  e_{\hat 2}= \left(\frac{1}{\Sigma}\right)^{1/2}\,\partial_\theta\,,\\
e_{\hat 3} &=&\frac{- 2\,M\,a\,r\,\sin \theta}{\left[\Delta\,\Sigma\,(\Sigma- 2\,M\,r)\right]^{1/2}}\,  \partial_t+ 
\left(\frac{\Sigma- 2\,M\,r}{\Delta\,\Sigma }\right)^{1/2}\,\frac{1}{\sin \theta}\,\partial_\phi\,.
\end{eqnarray}
This tetrad frame can be adapted with $e_{\hat 0}= \bar{u}$ to the reference observers that form a congruence of accelerated, nonexpanding and locally rotating world lines. The lack of expansion of the congruence is due to the alignment of its 4-velocity vector field with the timelike Killing direction of the exterior Kerr spacetime. 

The 4-acceleration of the fiducial observers is given by 
\begin{equation}\label{B7}
\mathbf{A}^\mu=\frac{De^{\mu}{}_{\hat 0}}{d\tau} = \Gamma^{\mu}_{\alpha \beta}\,e^{\alpha}{}_{\hat 0}\,e^{\beta}{}_{\hat 0}\,,
\end{equation}
where $\Gamma^{\mu}_{\alpha \beta}$ are the Kerr connection coefficients. Thus, 
\begin{equation}\label{B8}
\mathbf{A} = \frac{M\sqrt{\Delta}(r^2-a^2\cos^2\theta)}{\Sigma^{3/2}(\Delta -a^2 \sin^2\theta)}\,e_{\hat 1}
-\frac{2Mra^2 \sin \theta \cos \theta}{\Sigma^{3/2}(\Delta -a^2 \sin^2\theta)}\,e_{\hat 2}\,,
\end{equation}
where $\Delta -a^2 \sin^2\theta = \Sigma - 2\,M\,r$.
This acceleration counters the attraction of gravity; that is, it is due to forces that are not gravitational in origin and are necessary to keep the reference observer from falling into the source. 

To set up a nonrotating spatial frame along the world line of each fiducial observer, let $\boldsymbol{\Sigma}^\mu$ be a vector that is Fermi-Walker transported along  $e^{\mu}{}_{\hat 0}$; then,  
\begin{equation}\label{B9}
\frac{d\boldsymbol{\Sigma}^\mu}{d\tau}+\Gamma^{\mu}_{\alpha \beta}\, e^{\alpha}{}_{\hat 0}\,\boldsymbol{\Sigma}^{\beta}= (\mathbf{A}\cdot \boldsymbol{\Sigma})\,e^{\mu}{}_{\hat 0}-(e_{\hat 0} \cdot \boldsymbol{\Sigma})\,\mathbf{A}^\mu\,.
\end{equation}
We need a vector that  is orthogonal to the world line, namely, $e_{\hat 0} \cdot \boldsymbol{\Sigma} = 0$; furthermore, such a vector can be expressed in terms of the natural spatial frame $e^{\mu}{}_{\hat a}$. That is, $\boldsymbol{\Sigma}^\mu = s^{\hat a}\,e^{\mu}{}_{\hat a}$, which means via Eq.~\eqref{B9} that in the exterior Kerr spacetime
\begin{equation}\label{B10}
\frac{ds^{\hat 1}}{d\tau} = \frac{M\,a\,(r^2-a^2 \cos^2 \theta)\,\sin \theta}{\Sigma^{3/2}\,(\Sigma- 2\,M\,r)}\,s^{\hat 3}\,,
\end{equation}
\begin{equation}\label{B11}
\frac{ds^{\hat 2}}{d\tau} = - \frac{2\,M\,a\,r\,\sqrt{\Delta}\,\cos \theta}{\Sigma^{3/2}\,(\Sigma- 2\,M\,r)}\,s^{\hat 3}\,,
\end{equation}
\begin{equation}\label{B12}
\frac{ds^{\hat 3}}{d\tau} = \frac{M\,a}{\Sigma^{3/2}\,(\Sigma- 2\,M\,r)}\,[-(r^2-a^2 \cos^2 \theta)\,\sin \theta \,s^{\hat 1} + 2\,r \sqrt{\Delta}\,\cos \theta \, s^{\hat 2}]\,.
\end{equation}
These equations can be written in the form
\begin{equation}\label{B13}
\frac{ds^{\hat i}}{d\tau} =\epsilon^{\hat i \hat j \hat k}\,\Omega_{\hat j}\,s_{\hat k}\,, 
\end{equation}
where
\begin{equation}\label{B14}
\Omega^\mu = \Omega^{\hat a}\, e^{\mu}{}_{\hat a} = \beta (\cos \alpha \, e^{\mu}{}_{\hat 1} + \sin \alpha \, e^{\mu}{}_{\hat 2})
\end{equation}
is the proper precession vector. Here, the angle $\alpha$ and the proper precession frequency $\beta>0$ are defined by
\begin{equation}\label{B15}
\sin\alpha = \frac{(r^2-a^2\cos^2\theta)\,\sin \theta}{\mathbb{D}}\,, \qquad  \cos\alpha = \frac{2\,r\,\sqrt{\Delta} \,\cos\theta}{\mathbb{D}}\,
\end{equation}
and 
\begin{equation}\label{B16}
\beta = \frac{Ma}{\Sigma^{3/2}\,(\Sigma-2\,M\,r)}\,\mathbb{D}\,,
\end{equation}
where 
\begin{equation}\label{B17}
\mathbb{D} = \left[4\,r^2 \Delta\cos^2\theta + (r^2-a^2\cos^2\theta)^2\,\sin^2\theta\right]^{1/2}\,.
\end{equation}
It is interesting to note that $\alpha$ has the same range as the polar angle $\theta$, so that 
$\alpha:0\to \pi$ when $\theta: 0 \to \pi$.

Let us briefly digress here and use the results given above to correct a few typographical errors that occur in Ref.~\cite{Bini:2017uax}: In Eq.~(5) of~\cite{Bini:2017uax}, the temporal component of $e_{\hat 3}$ must be multiplied by $a$, the specific angular momentum of the Kerr source; moreover, in the denominators of Eqs.~(10)-(12) of~\cite{Bini:2017uax}, $(\Sigma - 2\, M\,r)$ must be replaced by $(\Sigma - 2\,M\,r)^{1/2}$. 

To construct a locally nonrotating spatial frame $\lambda^{\mu}{}_{\hat a}$ along $e^{\mu}{}_{\hat 0}$,  we can choose $\lambda^{\mu}{}_{\hat 1}$ to be the unit vector along $\Omega^\mu$; then, $\lambda^{\mu}{}_{\hat 2}$ and $\lambda^{\mu}{}_{\hat 3}$ are unit vectors in the plane orthogonal to $\lambda^{\mu}{}_{\hat 1}$ and precess with frequency $\beta$ about $\lambda^{\mu}{}_{\hat 1}$. Thus the second step involves the introduction of the orthonormal spatial frame $E^{\mu}{}_{\hat a}$,
\begin{eqnarray}\label{B18}
E_{\hat1} &=& \cos \alpha\, e_{\hat 1}+\sin \alpha\, e_{\hat 2}\,, \nonumber\\
E_{\hat 2} &=&-\sin \alpha\, e_{\hat 1}+\cos \alpha\, e_{\hat 2}\,, \nonumber\\
E_{\hat 3} &=& e_{\hat 3}\,.
\end{eqnarray}
Then in the final step the  Fermi-Walker transported triad $\lambda^{\mu}{}_{\hat a}$ is obtained from $E^{\mu}{}_{\hat a}$  by a simple rotation about $E^{\mu}{}_{\hat 1}$ with an angle of $\beta \tau$,
\begin{eqnarray}\label{B19}
\lambda_{\hat 1} &=& E_{\hat 1}\,,  \nonumber\\
\lambda_{\hat 2} &=&\cos (\beta \tau) \, E_{\hat 2}+\sin (\beta \tau) \, E_{\hat 3}\,, \nonumber\\
\lambda_{\hat 3} &=& -\sin (\beta \tau) \, E_{\hat 2}+\cos (\beta \tau) \, E_{\hat 3}\,.
\end{eqnarray}

Let us combine these rotation matrices in a new matrix $S$ defined by $\lambda^\mu{}_{\hat i}= S_{i}{}^j\, e^\mu{}_{\hat j}$,
where 
\beq\label{B20}
S= \left[
\begin{array}{ccc}
\cos \alpha & \sin\alpha & 0\cr
-\sin\alpha\,\cos\psi & \cos\alpha\,\cos\psi & \sin\psi\cr
\sin\alpha\,\sin\psi & -\cos\alpha\,\sin\psi & \cos \psi\cr
\end{array}
\right]\,.
\eeq
Here, $\psi = \beta\,\tau$. Moreover, the transpose of the rotation matrix $S$ is given by $S^T=S^{-1}$, 
\beq\label{B21}
S^T= \left[
\begin{array}{ccc}
\cos \alpha & -\sin\alpha\,\cos\psi  & \sin\alpha\,\sin\psi\cr
\sin\alpha & \cos\alpha\,\cos\psi & -\cos\alpha\,\sin\psi\cr
0 & \sin\psi & \cos \psi\cr
\end{array}
\right]\,,
\eeq
so that $e^\mu{}_{\hat i} = (S^T)_i{}^j\, \lambda^\mu{}_{\hat j}$. It is now possible to express the nongravitational acceleration of the fiducial observer in terms of the Fermi-Walker transported tetrad frame, namely,
\begin{equation}\label{B22}
\mathbf{A} = \frac{2Mr(r^2-a^2\cos^2\theta)\cos\theta}{\Sigma^{3/2}\,\mathbb{D}}\,\lambda_{\hat 1}
+\frac{M\sqrt{\Sigma\,\Delta}\sin\theta}{(\Sigma - 2 Mr)\mathbb{D}}\,(-\cos\psi \,\lambda_{\hat 2} + \sin\psi \,\lambda_{\hat 3})\,.
\end{equation}

This completes the construction of the adapted nonrotating tetrad frame  $\lambda^{\mu}{}_{\hat \alpha}$ along the world line of an arbitrary observer that is spatially at rest in the exterior Kerr spacetime; here, $\lambda^{\mu}{}_{\hat 0}=e^{\mu}{}_{\hat 0} = \bar{u}^\mu$. 

To establish a Fermi normal coordinate system along the world line of such a reference observer, it remains to calculate the spacetime curvature components given by Eq.~\eqref{A6}.

\section{Measured Components of Curvature}

In general, one can take into account the symmetries of the Riemann tensor and express Eq.~\eqref{A6} in the standard manner as a $6\times 6$ matrix $(R_{AB})$, where $A$ and $B$ are indices that belong to the set $\{01,02,03,23,31,12\}$. The general form of this matrix is 
\beq\label{C1}
\left[
\begin{array}{cc}
\mathcal {E} & \mathcal {H}\cr
\mathcal {H}^T & \mathcal {S}\cr 
\end{array}
\right]\,,
\eeq
where $\mathcal {E}$ and $\mathcal {S}$ are symmetric $3\times 3$ matrices and $\mathcal{H}$ is traceless. Here, the measured gravitoelectric components of the Riemann curvature tensor are represented by the relativistic tidal matrix $\mathcal{E}$. Similarly, $\mathcal{H}$ and  $\mathcal{S}$ represent its gravitomagnetic and spatial components, respectively. In the vacuum region exterior to material sources and free of nongravitational fields, the spacetime is Ricci flat as a consequence of Einstein's gravitational field equations and Eq.~\eqref{C1} becomes
\beq\label{C2}
\left[
\begin{array}{cc}
{\mathcal E} & {\mathcal H}\cr
{\mathcal H} & -{\mathcal E}\cr 
\end{array}
\right]\,,
\eeq
where  $ \mathcal{E}$ and $\mathcal {H}$ are symmetric and traceless. That is,  the Riemann curvature tensor degenerates in the Ricci flat case into the Weyl conformal curvature tensor whose gravitoelectric and gravitomagnetic components are then
\begin{equation}\label{C3} 
\mathcal{E}_{\hat a \hat b}= C_{\alpha\beta\gamma\delta}\,\lambda^\alpha{}_{\hat 0}\,\lambda^\beta{}_{\hat a}\, \lambda^\gamma{}_{\hat 0}\, \lambda^\delta{}_{\hat b}\,,\qquad \mathcal{H}_{\hat a \hat b}=C^*_{\alpha\beta\gamma\delta}\,\lambda^\alpha{}_{\hat 0}\,\lambda^\beta{}_{\hat a}\, \lambda^\gamma{}_{\hat 0}\, \lambda^\delta{}_{\hat b}\,,
\end{equation}
where  $C^*_{\alpha\beta\gamma\delta}$ is the unique dual of the Weyl tensor given by
\begin{equation}\label{C4} 
C^*_{\alpha\beta\gamma\delta}=\frac12 \eta^{\mu\nu}{}_{\alpha \beta}\,C_{\mu\nu\gamma\delta}\,,
\end{equation}
since the right and left duals of the Weyl  tensor coincide. Here, $\eta_{\mu\nu\rho\sigma}$ is the Levi-Civita tensor and in our convention, $\eta_{\hat 0 \hat 1 \hat 2 \hat 3}=1$, while $\eta_{\hat 0 \hat a \hat b \hat c}=\epsilon_{\hat a \hat b \hat c}$. Let us note that 
\begin{equation}\label{C5}
\mathcal{H}_{\hat a \hat b}= \frac12 \eta^{\mu\nu}{}_{\alpha \beta}\,C_{\mu\nu\gamma\delta}\,\lambda^\alpha{}_{\hat 0}\,\lambda^\beta{}_{\hat b}\, \lambda^\gamma{}_{\hat 0}\, \lambda^\delta{}_{\hat a}=\frac12 \eta^{\mu\nu}{}_{\hat 0 \hat b}\, C_{\mu\nu\hat 0 \hat a} = \frac12 C_{\hat 0 \hat a\hat c\hat d}\,\epsilon^{\hat c \hat d}{}_{\hat b}\,.
\end{equation}

We now turn to the explicit computation of curvature components as measured by observers at rest in the exterior Kerr spacetime. 

\subsection{Curvature of Kerr Spacetime as Measured by Observers at Rest}

Projected onto the canonical Petrov tetrad of the Kerr field~\cite{cart}, the Weyl tensor takes the form
\beq\label{C6}
\mathcal{E}_0 = \mathbb{E}\, \left[
\begin{array}{ccc}
-2 & 0 & 0\cr
0 & 1 & 0\cr
0 & 0 & 1\cr
\end{array}
\right]\,,\qquad
\mathcal{H}_0 = \mathbb{H}\, \left[
\begin{array}{ccc}
-2 & 0 & 0\cr
0 & 1 & 0\cr
0 & 0 & 1\cr
\end{array}
\right]\,,
\eeq
where,
\beq\label{C7}
\mathbb{E}+i\,\mathbb{H} = \frac{M}{(r+i\,a\,\cos \theta)^3}\,,
\eeq
so that
\beq\label{C8}
 \mathbb{E} = \frac{M r (r^2-3 a^2\cos^2 \theta)}{\Sigma^3}\,, \qquad 
 \mathbb{H}= -\frac{M a (3r^2- a^2\cos^2 \theta)\,\cos \theta }{\Sigma^3}\,.
\eeq 
 The evident ``parallelism"
between the gravitoelectric and gravitomagnetic components of curvature in Eq.~\eqref{C6} has to do with the degenerate nature of the Kerr field; in fact, it is of  type D in the Petrov classification.   We are  interested in the curvature of Kerr spacetime as measured by the family of observers at rest.   Projected onto the natural frame $e^{\mu}{}_{\hat \alpha}$ with axes that are primarily along the  Boyer-Lindquist coordinate directions, we find $(\mathcal {E'}, \mathcal {H'})$. Specifically, the nonvanishing components of the tidal matrix can be obtained 
from~\cite{Bini:2016xqg} 
\begin{eqnarray}\label{C9}
\mathcal {E'}_{\hat 1 \hat 1}&=& -2 \mathbb{E} \,\frac{\Delta +\frac{1}{2}\,a^2 \sin^2 \theta}{\Delta -a^2\sin^2 \theta}\,,\nonumber\\
\mathcal {E'}_{\hat 1 \hat 2}&=& -3\, a\,\sin \theta\, \mathbb{H} \,\frac{\Delta^{1/2}}{\Delta-a^2\,\sin^2 \theta}\,,\nonumber\\
\mathcal {E'}_{\hat 2 \hat 2}&=&  \mathbb{E} \,\frac{\Delta + 2\,a^2 \sin^2 \theta}{\Delta -a^2\sin^2 \theta}\,,\nonumber\\
{\mathcal E'}_{\hat 3 \hat 3}&=& \mathbb{E}\,.
\end{eqnarray}
Furthermore, the nonzero elements of the gravitomagnetic part of the Weyl curvature can be obtained from~\cite{Bini:2016xqg} 
\begin{eqnarray}\label{C10}
\mathcal {H'}_{\hat 1 \hat 1}&=& -2 \mathbb{H} \,\frac{\Delta +\frac{1}{2}\,a^2 \sin^2 \theta}{\Delta -a^2\sin^2 \theta}\,,\nonumber\\
\mathcal {H'}_{\hat 1 \hat 2}&=& 3\, a\,\sin \theta \,\mathbb{E} \,\frac{\Delta^{1/2}}{\Delta-a^2\,\sin^2 \theta}\,,\nonumber\\
\mathcal {H'}_{\hat 2 \hat 2}&=&  \mathbb{H} \,\frac{\Delta + 2\,a^2 \sin^2 \theta}{\Delta -a^2\sin^2 \theta}\,,\nonumber\\
{\mathcal H'}_{\hat 3 \hat 3}&=& \mathbb{H}\,.
\end{eqnarray}

We are actually interested in the measured components of curvature along the Fermi-Walker transported tetrad frame $\lambda^\mu{}_{\hat \alpha}$.  In this case, the measured components of the curvature tensor are given by Eq.~\eqref{C2}, where $(\mathcal{E}, \mathcal{H})$ are related to  $(\mathcal{E'}, \mathcal{H'})$  via a rotation $S$ given by Eq.~\eqref{B20}. Under such a rotation, it is straightforward to show that the gravitoelectric part (i.e., the relativistic  tidal matrix) and the gravitomagnetic part of the Weyl tensor undergo a similarity transformation, namely,
\begin{equation}\label{C11}
\mathcal{E}=S\,\mathcal{E'}\,S^{-1}\,, \qquad \mathcal{H}=S\,\mathcal{H'}\,S^{-1}\,.
\end{equation}
To express the measured curvature components explicitly, it proves convenient to define $\mathbb{P}$ and $\mathbb{Q}$ such that 
\begin{equation}\label{C12}
\mathbb{P} := \frac{1}{2}\,(\mathcal{E'}_{\hat 1 \hat 1}-\mathcal{E'}_{\hat 2 \hat 2})\,\cos 2\alpha +\mathcal{E'}_{\hat 1 \hat 2}\,\sin 2\alpha\,, \quad \mathbb{Q} := -\frac{1}{2}\,(\mathcal{E'}_{\hat 1 \hat 1}-\mathcal{E'}_{\hat 2 \hat 2})\,\sin 2\alpha +\mathcal{E'}_{\hat 1 \hat 2}\,\cos 2\alpha\,.
\end{equation}
Then, with $\psi = \beta\,\tau$, we have 
\begin{equation}\label{C13}
\mathcal{E}_{\hat 1 \hat 1} = -\frac{1}{2}\,\mathcal{E'}_{\hat 3 \hat 3}+ \mathbb{P}\,, \qquad \mathcal{E}_{\hat 1 \hat 2} = \mathbb{Q} \,\cos\psi\,,\qquad \mathcal{E}_{\hat 1 \hat 3} = -\mathbb{Q}\,\sin\psi\,,
\end{equation}
\begin{equation}\label{C14}
\mathcal{E}_{\hat 2 \hat 2} = -\left(\frac{1}{2}\,\mathcal{E'}_{\hat 3 \hat 3}+ \mathbb{P}\right)\cos^2\psi+\mathcal{E'}_{\hat 3 \hat 3}\,\sin^2\psi\,, \qquad \mathcal{E}_{\hat 2 \hat 3} = \left(\frac{3}{2}\,\mathcal{E'}_{\hat 3 \hat 3}+ \mathbb{P}\right)\sin\psi \cos\psi\,
\end{equation}
and
\begin{equation}\label{C15}
\mathcal{E}_{\hat 2 \hat 1}  = \mathcal{E}_{\hat 1 \hat 2}\,, \quad \mathcal{E}_{\hat 3 \hat 1}  = \mathcal{E}_{\hat 1 \hat 3}\,, \quad  \mathcal{E}_{\hat 3 \hat 2}  = \mathcal{E}_{\hat 2 \hat 3}\,, \quad \mathcal{E}_{\hat 3 \hat 3} = -\mathcal{E}_{\hat 1 \hat 1} -\mathcal{E}_{\hat 2 \hat 2}\,,
\end{equation}
since this matrix is symmetric and traceless; moreover, we have exactly the same type of expressions for $\mathcal{H}$. From 
$(\mathcal{E}, \mathcal{H})$,  we find  $R_{\hat \alpha \hat  \beta \hat \gamma \hat \delta}(T)$ employed in the Fermi normal coordinate system.

\section*{Acknowledgments}

I wish to thank C. Chicone and D. Bini for their past collaborations on the subject of the present paper.

\end{document}